\documentclass{article}
\usepackage{amssymb,cite,multirow} 
\def\1ad{\mbox{\normalsize $^1$}}
\def\2ad{\mbox{\normalsize $^2$}}
\def\3ad{\mbox{\normalsize $^3$}}
\def\4ad{\mbox{\normalsize $^4$}}
\def\5ad{\mbox{\normalsize $^5$}}
\def\6ad{\mbox{\normalsize $^6$}}
\def\7ad{\mbox{\normalsize $^7$}}
\def\8ad{\mbox{\normalsize $^8$}}

\makeatletter
\@addtoreset{equation}{section}
\makeatother

\def\beq{\begin{equation}}                     %
\def\eeq{\end{equation}}                       %
\def\bea{\begin{eqnarray}}                     
\def\eea{\end{eqnarray}}                       
\def\nn{\nonumber} 
 
\def\cy {Calabi--Yau}

\def\clss {Clifford$(6,6)$} 
\def\eij {e^{iJ}}
\def\stt {SU(3)$\times$SU(3)}
 
\def\0 {\nonumber}

\def\Ox{\Omega}

\newcommand{\sla}{\slash\!\!\!\!}

\def\N{\mathcal N}
\def\Phin{\hat \Phi}

\def\tts{$T \oplus T^*$ }

\newcommand{\bbR}{\mathbb{R}}
\newcommand{\bbC}{\mathbb{C}}

\begin{document}
\setcounter{page}{0}
\begin{titlepage}
\titlepage
\rightline{hep-th/0505212}
\rightline{CPHT-RR 029.0505}
\rightline{SPhT-T05/079}
\rightline{LPTENS-05/16}
\rightline{SU-ITP-05/18}
\vskip 3cm
\centerline{{ \bf \Large Generalized structures of $\N=1$ vacua}}
\vskip 1.5cm
\centerline{Mariana Gra{\~n}a$^{a,b}$, Ruben
Minasian$^{c,a}$, Michela Petrini$^{a,d}$
and Alessandro Tomasiello$^e$}
\begin{center}
\em $^a$Centre de Physique Th{\'e}orique, Ecole
Polytechnique
\\91128 Palaiseau Cedex, France\\
\vskip .4cm
$^b$Laboratoire de Physique Th{\'e}orique, Ecole Normale
Sup{\'e}rieure\\
24, Rue Lhomond 75231 Paris Cedex 05, France\\
\vskip .4cm
$^c$ CEA/DSM/SPhT, Unit{\'e} de recherche associ{\'e}e
au CNRS, CEA/Saclay \\
 91191 Gif-sur-Yvette, France\\
\vskip .4cm
$^d$Laboratoire de Math{\'e}matiques et Physique Th{\'e}orique,
 Universit{\'e} Fran{\c c}ois Rabelais\\ 
Parc de Grandmont 37200, Tours, France
\vskip .4cm
$^e$ITP, Stanford University, Stanford CA 94305-4060
\end{center}
\vskip 1.5cm  
\begin{abstract}

We characterize ${\cal N}=1$ vacua of type II theories 
in terms of generalized complex structure on the internal manifold $M$.
The structure group of $T(M) \oplus T^*(M)$ being \stt\ implies the
existence of two pure spinors $\Phi_1$ and $\Phi_2$. The conditions for preserving 
$\N=1$ supersymmetry turn out to be simple generalizations of equations that
have appeared in the context of ${\cal N}=2$ and topological strings.
They are  $(d+ H\wedge)\Phi_1=0$ and  
$(d+H\wedge)\Phi_2=F_{\mathrm{RR}}$. 
The equation for the first pure spinor implies 
that the internal space is a twisted generalized Calabi-Yau manifold of a hybrid
complex-symplectic type, while the RR-fields serve
as an integrability defect for the second.

\end{abstract}

\vfill
\begin{flushleft}
{\today}\\
\end{flushleft}
\end{titlepage}

\newpage
\large

\section{Introduction}

In compactifications of type II theories to four dimensions, requiring 
unbroken ${\cal N}=2$ supersymmetry in the absence of fluxes 
leads to a simple and well--studied condition: 
the six--dimensional internal manifold should be Calabi--Yau. In recent years, 
there have been many attempts to find a similarly sleek answer in the presence
of fluxes, both for ${\cal N}=2$ and ${\cal N}=1$ supersymmetries in four 
dimensions. 

Calabi-Yau manifolds admit a covariantly constant spinor. This condition 
has two parts, an algebraic one, namely 
the existence of a globally defined non-vanishing spinor; and a differential
one, the requirement that it be covariantly constant. The first requirement
cannot be relaxed if we want to have in the  four--dimensional theory
eight well--defined supercharges, or in other words an $\N=2$ effective action. 
Vacua with  spontaneously broken supercharges or four--dimensional vacua 
other than Minkowski require relaxing the condition of the spinor to be 
covariantly constant.

Thus the  existence of  supersymmetric vacua of four--dimensional
$\N=2$ effective theories in compactifications of type II
theories in presence of NS and RR fluxes imposes a number of algebraic 
and differential conditions on the internal space. This paper is devoted
to analysis of these conditions. 
The algebraic part says that the tangent plus cotangent bundle
of the internal manifold, $T \oplus T^*$, must have structure group \stt \cite{hitchin,gualtieri,jw} (see also \cite{Witt} for the case of $G_2 \times G_2$ 
structures). 
This is the generic condition ensuring $\N=2$ supersymmetry of the  
effective action in four-dimensions \cite{GLW} and implies 
in particular the existence of two nowhere vanishing globally defined 
\clss\ pure spinors
Both can be represented just as sums of usual
differential forms  of even or odd rank over the manifold, and are 
denoted as $\Phi_1$ and $\Phi_2$. The differential part then 
says that the two pure spinors should satisfy
\begin{equation}
  \label{eq:summa}
  e^{-2A+\phi}(d+H\wedge)(e^{2A-\phi} \Phi_1)=0\ ,\qquad
  e^{-2A+\phi}(d+H\wedge)(e^{2A-\phi} \Phi_2)=dA\wedge \bar\Phi_2 + F\ .
\end{equation}
Here,
$e^{2A}$ is the warp factor, $\phi$ the dilaton, 
$H$ the NS three--form and
$F$ a weighted sum of the RR fields, 
see eqs.(\ref{a+}--\ref{b-}).
$\Phi_1$ has the same parity as the RR fields, namely even in IIA and odd in IIB.
The equations above together with the equations for the norm of the pure spinors,
(\ref{eq:norm}), 
give the necessary and sufficient conditions on the manifold
and the fluxes to have unbroken $\N=1$ supersymmetry on warped Minkowski 
space\footnote{In order
to satisfy all the equations of motion, we have to impose additionally Bianchi 
identities and the equations of motion for the fluxes (see for example
\cite{Tsimpis}).}. 
Equations  (\ref{a+}--\ref{b-}) also take into
account the possibility of a nonzero cosmological constant in four dimensions. 
 
The normalizations of the pure spinors are such that these equations 
are not applicable (or rather, do not give any information, as the 
norms are zero in that case) to the case in which RR fields
are absent. However the supersymmetry conditions for zero RR fluxes are  
very well known. As we will 
review later, the NS flux in that case does not enter in the 
equations as a simple twist of  the exterior derivative, as it does in (\ref{eq:summa})
(or in more complete form in (\ref{a+}--\ref{b-})).

Besides being compact, these equations fit very well in the mathematical 
framework of generalized complex geometry \cite{hitchin,gualtieri}. In 
particular, the first equation in (\ref{eq:summa}) implies that the manifold
must be twisted generalized Calabi--Yau. This fact has been noted already in 
\cite{gmpt} for cases in which $T$ has  
SU(3) structure. The SU(3) structure vacua are however just special
cases of the  more general \stt\  on $T\oplus T^*$ considered here. 
The SU(3) structure vacua correspond to  either 
complex (and with vanishing $c_1$) or symplectic manifolds, which are the two 
particular cases which inspired the definition of generalized Calabi--Yau.
In the generic \stt\ case, vacua can be a complex--symplectic hybrid,
namely a manifold that is locally a product of a complex $k$-fold times an
$6-2k$ symplectic manifold.  

Physically, the generalized Calabi--Yau condition has also been argued to 
imply the existence of a topological model \cite{kapustin,zucchini},
not necessarily coming from the twisting of a $(2,2)$ model,
which generalizes
the A and B models. In other words, we find that all ${\cal N}=1$ Minkowski 
vacua  have
an underlying topological model. When there is a $(2,2)$ model, both 
pure spinors  are closed under $(d+H\wedge)$ \cite{gualtieri}, 
which reflects the 
fact that two topological models can be defined. 
This condition unifies the \cy\ case and the $(2,2)$ models
of \cite{ghr}. It corresponds to an unbroken ${\cal N}=2$  
in the target space,  and it has recently  
been found from supergravity in \cite{jw}. Although the $\N =2$ 
requirement 
 of having two twisted closed pure spinors looks like our
$\N=1$ equations (\ref{eq:summa}) for $F=A=0$, we stress again that  
(\ref{eq:summa}) applies  only when the RR fluxes are non zero. 
Therefore we cannot obtain from there the $\N =1$ equations for pure NS 
flux, which correspond to a $(2,1)$ model.

Another feature of (\ref{eq:summa}) is that they are essentially identical
for IIA and IIB. This suggests there must exist some form of mirror symmetry
for these compactifications exchanging the even and odd pure spinors \cite{fmt,pg,t}. 
As far as we know, mirror symmetry could 
even be present when supersymmetry is spontaneously broken 
\cite{glmw,fmt,t}. For the case at hand of unbroken ${\cal N}=1$, however, 
this is 
made particularly concrete by the remark above that all vacua have an 
underlying topological model; mirror symmetry has long been viewed 
\cite{kontsevich} as an exchange of topological models, without necessarily
involving \cy's. 

Presumably there are connections to more recent lines of thought relating
Hitchin functionals \cite{hf} to topological theories \cite{dgnv,gv,n}.
Particularly promising seems the results in \cite{pw} about the quantization
of the functional, which relate directly to generalized \cy's with \stt\ structure.

On the practical side
\stt\ structures allow to treat more easily cases which would
be otherwise complicated, and that for this reason may have been regarded
so far as pathological or unpractical. Most mathematical approaches, starting 
from \cite{gmpw},
have to date focused on structure groups on $T$. 
For SU(3) structure the theory of intrinsic torsions involved 
is useful and manageable; already for SU(2), which
is the intersection of two SU(3) structures, it becomes less tractable,
as more representations appear in the game. 
Furthermore, there are cases in between SU(3) and SU(2)
which do not really deserve a name as a structure on $T$, such as
two SU(3) structures degenerating at some points. 
All these cases can be treated on the same footing if we consider 
structures on $T \oplus T^*$: they are \stt\ structures.
The pure SU(3) structure case is rather
the exception than the rule;
with \stt\ structures more cases are available. 
An extreme example can be found in \cite{cg}, where
the structure is actually the trivial group: some nilmanifolds are
complex, some are symplectic, most are neither; but they all are generalized 
complex.

In the following sections we give a detailed explanation of the conditions to have $\N=1$ 
supersymmetry.
In section \ref{sec:algebraic} we review the algebraic conditions, 
describing the SU(3), SU(2) and generic \stt\ cases. We  
give the differential conditions in section \ref{sec:diff}, and discuss their implications.
Section \ref{sec:app} contains a discussion of the connections to other results and applications.
Then in appendix \ref{app:top} we briefly review the relations between the pure spinor equations and
topological models. We also propose, in appendix \ref{app:N1}, 
 how to interpret the $\N=1$ condition for vacua
without RR fluxes in terms of Courant brackets.

\section{The algebraic part: structure groups} \label{sec:algebraic}

In this section we discuss how \stt\ structures on $T \oplus T^*$ 
describe in a unified way the various structures arising on $T$
\cite{hitchin,gualtieri,jw}. The
real advantages of the \stt\ description will however emerge in the
discussion of the differential conditions. 

The supersymmetry transformations for type II theories contain two 
ten--dimensio- nal Majorana-Weyl spinor parameters $\epsilon_{1,2}$. 
If the ten--dimensional manifold is topologically a homogeneous 
four--dimensional 
space (AdS, Minkowski) 
times an internal six--manifold, the ten-dimensional
spinors can be decomposed into spinors in four dimensions
times internal spinors. Since we are interested in backgrounds preserving
four-dimensional ${\cal N}=1$ supersymmetry, there should be a single
four dimensional conserved spinor. We therefore write
\begin{equation}
  \label{eq:spindec}
  \begin{array}{ccc}
\epsilon^1 & = &  \zeta_{+} \otimes \eta_{+}^1 + \zeta_{-} \otimes 
\eta_{-}^1 \, ,
\nonumber\\ 
\epsilon^2 & = & \zeta_{+} \otimes \eta_{-}^2  
+ \zeta_{-} \otimes \eta_{+}^2 \, ,
  \end{array}\quad \mathrm{(IIA);} \quad \quad
\epsilon^i = \zeta_{+} \otimes \eta_{+}^i + \zeta_{-} \otimes \eta^{i}_{-} \, ,
\quad \mathrm{(IIB)} \, ,
\end{equation}
for any four--dimensional spinor $\zeta_+$, with $\zeta_-$ being
its Majorana conjugate\footnote{We could imagine a
more general way of relating the two four--dimensional spinors in 
$\epsilon_{1,2}$, like for example $\xi_1= A_{\mu \nu} 
\gamma^{\mu \nu} \xi_2$, but maximal symmetry in
four dimensions is only compatible with (\ref{eq:spindec}); see for example \cite{ls}.}. 
Also, $(\eta^{i}_{+})^* = \eta^i_-$, in such 
a way that $\epsilon^i$ are Majorana in ten dimensions. 
Here, we have not yet taken the spinors to be normalized
to any particular value. 


Given two spinors on the internal manifold, there are different
possible relations among them, that lead to different structures.

To begin with, the spinors $\eta^1$ and $\eta^2$
 may simply be proportional. Then they
define what is called an {\it SU(3) structure} 
(this is uniquely defined by a nowhere vanishing spinor invariant 
under SU(3) $\subset $ SO(6), not necessarily covariantly constant). 
A prominent example of an SU(3) structure manifold is a \cy\ 3-fold.
In that case, the invariant spinor is also covariantly constant
and the structure group coincides with the holonomy group. 
The case of SU(3) structure is particularly simple
-- and hence much studied -- due to the fact that few representations
 are involved.
The manifold can be characterized either by the SU(3) invariant
spinor, or by an SU(3) invariant 
real two-form $J$ and a complex three-form $\Omega$, obeying 
$J\wedge\Omega=0$, $i\Omega\wedge\bar\Omega=\frac{4}{3}J^3$.
However for several purposes it is better to deal with  $\eij$  rather 
with $J$ alone. 
Already in the study of branes 
on \cy\ manifolds, $\eij$ emerges as the mirror of $\Omega$. This feature 
persists, at least at local level, for general SU(3) structure \cite{fmt}. 
In \cite{gmpt}, it was also shown that the ${\cal N}=1$ condition in 
supergravity  naturally
uses $\eij$. We will come back to this issue in the next subsection.

The other extreme choice for the relation between the spinors $\eta^1$ and $\eta^2$ 
is to take them never parallel.  In this case  there is a bilinear which
defines a complex vector field without zeros
\beq \label{vector}
\eta^{1\, \dagger}_{+} \gamma_m \eta^2_- = v_m - iw_m \, .
\eeq  
The  
two spinors give rise to two different SU(3) structures whose 
intersection gives an {\it SU(2) structure}. Each  SU(3) structure has 
an associated almost complex structure $J_{m}{}^p$.\footnote{For
a given SU(3) structure, we call the two--form $J$ and the almost complex 
structure $J_m{}^n$ induced by $\Omega$ with the same name. This should not 
lead to a confusion, for the complex structure the indexes will always be 
written explicitly.} The product of
the two almost complex structures, $J_{1m}{}^p J_{2p}{}^n$, 
is a tensor which squares to 1 and has four negative and two positive 
eigenvalues; it is called an {\it almost product 
structure} and can
be used to split the tangent space at every point (and hence also all the 
bundles $\Lambda^p T^*$) in four plus two dimensions\footnote{
\label{foot:productstructure}
$j$ and $\omega$ can be thought of as an SU(2) structure in four dimensions, 
and $v$ and $w$ as a trivial structure in two. 
This split however does not mean a priori that there is any four--dimensional 
submanifold along which $j$ and $\omega$ are defined; this would mean the
almost product structure is  integrable.}:
\begin{equation}
  \label{eq:su2}
  J_{1,2}= j \pm v\wedge w\ , \qquad \Omega_{1,2}=\omega\wedge(v \pm i w)\, .
\end{equation}
The forms $j$, $\omega$,  $v$ and $w$ define the SU(2) structure 
in 6 dimensions. They are all nowhere vanishing. 
Alternatively, it is also possible to define the  SU(2) structure from 
an SU(3) one, say $J_1$ and $\Omega_1$, and one vector $v$.

Both SU(3) and SU(2) impose some topological constraints on the manifold. 
The one imposed by SU(2) is stronger: due to the fact
that the two spinors are never parallel, the vector $v$ 
in (\ref{eq:su2}) is nowhere vanishing.
This is of course possible if and only if the Euler characteristic 
$\chi$ of the manifold vanishes. Thus the condition
$\chi(M)=0$ ensures the existence of a topological SU(2) structure on
six-manifolds with SU(3)  structure.
On the contrary when the two spinors are parallel the vector
defined in (\ref{vector}) vanishes 
everywhere.

There are  however more general situations where the two spinors $\eta^1$ 
and $\eta^2$ can become parallel at points on the manifolds and this does not
impose any extra topological constraint with respect to SU(3) structure.
To treat all these
cases in a uniform way it is better to consider structures on the sum of the 
tangent and cotangent bundles rather than on the tangent bundle alone.

\subsection{Structures on \tts}

Enlarging the space by combining the tangent
and cotangent bundles in a single bundle, $T \oplus T^*$, 
allows not only  
to give a unified description of the structures on T we mentioned above, but
also to  understand the 
mathematical meaning of the formal sum of forms $\eij$ and what $\Omega$ and $\eij$ have in 
common. 
 
In general, a formal sum of forms can be viewed as a 
representation of O$(6,6)$, which is the structure group 
of $T \oplus T^*$. In fact, $\Omega$ and $\eij$  transform under  O$(6,6)$ 
in exactly the same fashion as the formal sum of RR fields transforms
under the T--duality group. Moreover, they share another very important
property -- {\it purity} -- which will be explained shortly.

Let us first consider the spinor group and the representation of the Clifford
algebra corresponding to O$(6,6)$, which is called \clss. It is defined by 
$$
\{ \lambda^m, \lambda^n\} =0\ , \qquad 
\{ \lambda^m, \iota_n\} = \delta^m{}_n \ , \qquad
\{ \iota_m, \iota_n \}=0\  \, ,
$$
where $\delta^m{}_n$ is the 6+6--dimensional metric 
${{0 \ 1 }\choose {1 \ 0}}$ on \tts. 
A useful representation is given by 
the six wedge operators and the six contractions 
\beq
\lambda^m\equiv dx^m \wedge \qquad \quad \iota_n\equiv\iota_{\partial_n} \, .
\eeq 
Starting from a Clifford vacuum one can generate any form by acting 
with the appropriate  $\lambda$'s and $\iota$'s. It is then clear
why a formal sum of forms in $\Lambda^{\bullet} T^*$ is a \clss\ spinor. 
There are irreducible (``Majorana-Weyl'') representations 
of Spin(6,6), which correspond to real (``Majorana'')
even or odd forms (``Weyl''). Here we will
work with Weyl Clifford(6,6) spinors $\Phi_{\pm} \in \Lambda^{even/odd} T^*$.

A Clifford(6,6) spinor 
is pure if there exist six linear combinations of the $\{ \lambda^m, 
\iota_n \}$ which annihilate it.

We argued that a sum of forms is a \clss\ spinor. It can 
nevertheless also be mapped to a 
bispinor, using the Clifford map:
\begin{equation}
  \label{eq:clmap}
C\equiv\sum_k \frac{1}{k!}C^{(k)}_{i_1\ldots i_k} dx^{i_i}\wedge\ldots\wedge dx^{i_k}\qquad
\longleftrightarrow\qquad
\sla C \equiv
\sum_k \frac{1}{k!}C^{(k)}_{i_1\ldots i_k} \gamma^{i_i\ldots i_k}_{\alpha\beta} \ .
\end{equation}
On a bispinor, we can act with usual gamma matrices from the left 
(which we denote as $ \buildrel\to\over{\gamma^m}$)  and from the right 
(which we denote as $ \buildrel\leftarrow\over{\gamma^m}$)
\begin{equation}
  \label{eq:gammamap}
 \buildrel\to\over{\gamma^m} = \frac12(\lambda^m+g^{mn}\iota_n)\ , \qquad
  \buildrel\leftarrow\over{\gamma^m} = \frac12(\lambda^m\pm g^{mn}\iota_n)\ \, ,
\end{equation}
where the $\pm$ sign is the parity of the spinor on which 
$ \buildrel\leftarrow\over{\gamma^m}$ acts.
The result of this action
is still a bispinor. This gives a map between 
two copies of Clifford(6) and \clss. This map was used extensively 
in \cite{gmpt} and independently in \cite{jw,Witt}.

One can use the Clifford map (\ref{eq:clmap}) to determine  the 
pure spinors on $T \oplus T^*$, $\Phi_+$ and $\Phi_-$, as the counter-image of bispinors,
namely
\begin{equation}
  \label{eq:genpure}
 \Phi_+\equiv \eta^1_+\otimes \eta_{+}^{2 \, \dagger}\ , \qquad 
 \Phi_-\equiv \eta^1_+\otimes \eta_{-}^{2 \, \dagger}\ .
\end{equation}
The bispinor picture is useful to check 
that these spinors are pure.
The six linear combinations of 
the $\{ \lambda^m, \iota_n\}$ required to show purity   
of both $\Phi_{\pm}$  are the counter-image
under (\ref{eq:gammamap}) of ordinary gamma matrices, three acting from the left
and three from the right\footnote{The fact that each of the spinors $\eta_+^i$ has
three annihilators among the six gamma matrices can be 
rephrased in terms of purity for Clifford(6) spinors: all spinors in six dimensions
are pure (see for example \cite{lm}).} :
\begin{equation}
  \label{eq:genann}
(\delta +iJ_1)_m{}^n \gamma_n \,  \eta^1_+\otimes \eta_{\pm}^{2 \, \dagger}=0 \ ,
\qquad  
   \eta^1_+\otimes \eta_{\pm}^{2 \, \dagger}\,\gamma_n 
(\delta \mp iJ_2)_m{}^n =0 \ .
\end{equation}

The purity condition can also be reformulated in terms of structure groups on 
$T \oplus T^*$: each pure spinor reduces the structure
group from O$(6,6)$ to SU$(3,3)$.  Furthermore, if two given pure spinors
satisfy a compatibility condition, namely they
have exactly three common annihilators, the structure is reduced to 
\stt.  
From eq. (\ref{eq:genann}) it is easy to see that 
$\eta^1_+\otimes \eta_{\pm}^{2 \, \dagger}$ are 
compatible: the three annihilators on the left are common to both pure 
spinors. $\Phi_{\pm}$ define therefore an \stt\ structure on \tts.

We will now look again at the particular cases of \stt\ structures 
introduced before, and give the corresponding Clifford(6,6) pure
spinors. 

\subsubsection{Pure spinors for SU(3) structure on $T$} \label{sec:SU3}

If the manifold has  SU(3) structure there are two pure spinors that one 
can build with the forms $J$ and $\Omega$. These are precisely $\eij$ and 
$\Omega$. In terms of the invariant spinors on the manifold we can define
these as special cases of (\ref{eq:genpure}) 
\begin{equation}
  \label{eq:fierz}
\Phin_+=
\frac{1}{8} 
e^{- i J}
\, , \qquad 
\Phin_-= 
-\frac{i}{8} {\Omega} \, .
\end{equation}
Differently from (\ref{eq:genpure}), 
the pure spinors $\Phin_{\pm}$ are normalized, 
that is
to say that they are built out of a single spinor 
$\eta=\eta_1 =\eta_2$ normalized to one; 
in the  future we shall reserve the hats for the normalized quantities.

From (\ref{eq:genann}) we can check that 
both $\Omega$ and $\eij$ are indeed pure
\begin{equation}
  \label{eq:annsu3}
\begin{array}{c}
\lambda^i \Omega=0\ ,\\ \iota_{\bar i}\Omega =0\ 
\end{array}
\ ; \qquad \qquad
(\iota_m - i J_{mn} \lambda^n) e^{-iJ}=0\ ,
\end{equation}
and that they are compatible:
the three common
annihilators are $\iota_{\bar i} - i J_{\bar i j}\lambda^j$. Hence the pair
$(\eij,\Omega)$ reduces the structure on \tts  to \stt.

\subsubsection{Pure spinors for SU(2) structure on $T$} \label{sec:SU2}

SU(2) structure on $T$ is also a particular 
case of \stt. $\eta^1$ and $\eta^2$ are never parallel.  
For example, when they are 
everywhere orthogonal, they define what is known as a static SU(2) 
structure. In this case,
the two compatible pure spinors of (\ref{eq:genann}) are  
\begin{equation}
  \label{eq:puresu2}
\Phin_+ = e^{-i\,j}\wedge(v+i w)\ , \qquad  \Phin_-=
\omega\wedge e^{-i\,v\wedge w} \ ,
\end{equation}
in terms of the almost product structure introduced above (\ref{eq:su2}).
Notice that both have a kind of four--two split: the first looks like $\eij$ 
in four dimensions and like $\Omega$ in two dimensions; the second like 
$\Omega$ in four dimensions and like $\eij$ in two dimensions.

\subsubsection{The generic \stt\ case on \tts} \label{sec:S33}

We have detailed so far two cases of \stt\ structures on $T \oplus T^*$: SU(3) on $T$ 
and SU(2) on $T$. In the first case, the two spinors are proportional; in the
second case, they are never parallel. As already mentioned, 
in the generic case they can be parallel at some points. 


For this generic case, let us first write the
spinor $\eta^2_+$ as a linear combination of elements in
a basis derived from $\eta^1$.
This basis for Clifford(6) spinors is the usual one obtained acting
on a Clifford vacuum $\eta^1_+$: $\eta^1_+, \gamma^m \eta^1_+, \gamma^m \eta^1_-, 
\eta^1_-$ (note that by purity of the spinor $\eta^1$, only 
three out of the six gamma matrices give a nonzero spinor).
To have positive chirality $\eta^2_+$ can only be a linear combination of
$\eta^1_+$ and $\gamma^m \eta^1_-$; one can write, in a similar notation as
in \cite{jw},
\begin{equation}
  \label{eq:spindep}
  \eta^2_+ = c \, \eta^1_+ + (v +iw)\cdot \eta^1_-
\end{equation}
where $(v+iw)\cdot$ is the Clifford action $(v+iw)_m \gamma^m$,
and $c$ is a complex function. Using this, 
equations (\ref{eq:fierz}) and the four--two split given in 
(\ref{eq:su2}), one finds\footnote{We are using that the SU(3)
structure defined by $\eta^1$ is given by $J=j+ v \wedge w$, 
$\Ox=\omega  \wedge (v+iw)$.} \cite{jw}
\begin{equation}
  \label{eq:genpureforms}
\Phin_+=\frac18(\bar c \, e^{-i\,j}-i  \omega)\wedge
e^{-iv\wedge w}\ , \qquad
\Phin_-=- \frac18( e^{-i\,j}+i c \, \omega)\wedge
(v+iw)\ .
\end{equation}
These formulae are written using the {\it local} SU(2) structure defined 
by the pair of 
spinors or equivalently the intersection of two SU(3) structures. 
The spinors in (\ref{eq:genpureforms}) are the intrinsic invariant 
objects of the \stt\ structure, while  the pure spinors defined by 
$\eta^i_{+} \otimes \eta^{i \, \dagger}_\pm$
are invariant under only one SU(3). No particular 
properties of $j$, $\omega$, $v$ and $w$ are assumed 
- the structure group of \tts 
being \stt\ simply tells us that $\Phin_+$ and $\Phin_-$ are well defined and nowhere 
vanishing\footnote{At the points where the two spinors $\eta^1, \eta^2$
become parallel, i.e. when $(v+i w)=0$, the lower case 
forms  $j$ and $\omega$ are not well defined, but
we should think of the products $e^{i(j+v \wedge w)}$ and $\omega \wedge
(v + i w)$  as $e^{iJ}$ and $\Ox$ (the two pure bispinors
 built out of $\eta^1$), which are well defined.}.  
If $v$ and $w$ turn out to be nowhere-vanishing, 
the internal space $M$ is more restricted than we have generally assumed
(namely 
besides admitting an almost complex structure and having a vanishing first 
Chern class, it should have $\chi=0$). 

Let us summarize the normalized pure spinors in the various cases:
\begin{center}
\renewcommand{\arraystretch}{1.5}
\begin{tabular}{c||c|c|}\cline{2-3}
& $\Phin_+$ &$\Phin_-$\\\hline\hline
\multicolumn{1}{|c||}{SU(3)}& $e^{-iJ}$ &$\Omega$\\\hline
\multicolumn{1}{|c||}{SU(2)}& $\omega\wedge e^{-i\,v\wedge w}$
&$e^{-i\,j}\wedge(v+i w)$\\\hline
\multicolumn{1}{|c||}{\stt}& $(\bar c \, e^{-i\,j}-i  \omega)\wedge
e^{-iv\wedge w}$ & $-( e^{-i\,j}+i c \, \omega)\wedge
(v+iw)$\\\hline
\end{tabular} 

\vskip 0.1cm

\small{{\bf  Table 1:} normalized pure spinors for the various structures}
\end{center}

We remind the reader that here we have presented the normalized pure 
spinors ($c$ and the norm of the complex vector
should be related by $|c|^2+ |v+iw|^2 =1$).
The differential equations for $\N=1$ supersymmetry given in the next section 
will be for their non-normalized counterparts.

The analysis of this section only regards topological properties of the 
manifold. Much more meaningful constraints are obtained when considering
the integrability properties of the structures considered so far. This is 
the topic of the next section.

\section{The differential part: integrability}
\label{sec:diff}

\stt\ structures describe in a unified way several types of
structures on $T$. In this section we will see that 
the conditions for ${\cal N}=1$ vacua translate into conditions for 
the integrability of an \stt\ structure.

We will first present the equations, then we give a brief
explanation of how they were obtained and finally discuss their 
mathematical interpretation. 

Preserved supersymmetry imposes differential equations
on the Clifford(6) spinors. As a consequence, the
pure Clifford(6,6) spinors,  given in (\ref{eq:genpure})  
as bispinors and in (\ref{eq:genpureforms}) as formal sums
of forms, have to obey certain differential conditions.
In order to preserve $\N=1$ supersymmetry, the conditions are
\bea
\label{a+}
e^{-2A+\phi}(d+H\wedge) (e^{2A-\phi}\Phi_+)&=&
2\mu\,e^{-A}\,\mathrm{Re}(\Phi_-) \, ,
\\
\label{a-}
e^{-2A+\phi}(d+H\wedge) (e^{2A-\phi}\Phi_-)&=&
3 i\,e^{-A} \,\mathrm{Im}(\bar \mu \Phi_+)
 +dA \wedge \bar\Phi_- \\
\nn&&-\frac 1{16}e^{\phi}\Big[ ( |a|^2 - |b|^2) F_{\mathrm{IIA}\,-}
 +i (|a|^2 + |b|^2)*F_{\mathrm{IIA}\,+}\Big]   \, ,
\eea  
for type IIA, and
\bea
\label{b+}
e^{-2A+\phi}(d-H\wedge) (e^{2A-\phi}\Phi_+)&=&
-3i\,e^{-A}\, \mathrm{Im}(\bar\mu  \Phi_-)+ dA\wedge\bar\Phi_+ \\
&& 
\label{b-}+\frac1{16} e^\phi
\Big[ (|a|^2 - |b|^2) F_{\mathrm{IIB}\,+} +i (|a|^2 + |b|^2)*F_{\mathrm{IIB}\,-}
\Big]
\ , \nonumber \\
e^{-2A+\phi}(d-H\wedge) (e^{2A-\phi}\Phi_-)&=&
-2\mu\,e^{-A}\,\mathrm{Re}(\Phi_+)  \, ,
\eea 
for type IIB. 
Let us explain the various pieces in these equations: in both theories
$F$ is a formal sum of modified RR fluxes (i.~e.~obeying 
a non standard Bianchi identity $dF_n= H  \wedge F_{n-3}$)
\footnote{All fluxes here are internal,
any piece along spacetime has been traded for a dual internal flux. See
the end of this section for more details.}$^,$\footnote{We thank L.~Martucci 
and P.~Smyth for correcting signs in an earlier version of   
(\ref{a-},\ref{b+}).} 
\begin{equation}
  \label{eq:F}
  F_{\mathrm{IIA}\,\pm}=F_0\pm F_2+F_4\pm F_6\ , \qquad 
F_{\mathrm{IIB}\, \pm}=F_1 \pm F_3+F_5\ .
\end{equation}
$A$ is the warp factor in a warped product metric of the form
\beq
ds^2= e^{2A} g_{\mu \nu} dx^\mu dx^\nu + ds_6^2
\eeq
where $g_{\mu \nu}$ is a maximally symmetric space with
zero or negative cosmological constant $\Lambda$ (Minkowski, or AdS). $\Lambda$ is 
related to the complex quantity $\mu$ 
by
\beq \label{cc}
\Lambda = - |\mu|^2 \, .
\eeq
Finally,
\beq \label{defnorms}
|a|^2=|\eta_1^+|^2 \, , \qquad |b|^2=|\eta_2^+|^2= 
|a|^2 (|c|^2 + |v+i w|^2) \, ,
\eeq  
where in the last equality we have used (\ref{eq:spindep}). 
From (\ref{eq:genpure}) we can see that
the norms of the pure spinors are also given in terms of
$a$ and $b$ by
\beq \label{normpure}
|\Phi_\pm|^2=|a|^2 |b|^2 \, .
\eeq
$\N=1$ supersymmetry imposes the following relation between the norms
\begin{equation}
  \label{eq:norm}
  d |a|^2 = |b|^2 d A\ , \qquad d|b|^2 = |a|^2 dA\ , 
\end{equation}
for both IIA and IIB.

A very important remark is in order: equations (\ref{a+}-\ref{b-}) do not
cover the case $F=0$, ${\cal N}=1$ vacua. 
From the Hodge decomposition (\ref{eq:hodge}),
it is easy to see that setting $F=0$ implies $dA =0$. Then (\ref{eq:norm}) tells that $a$ 
and $b$ have constant norms. If both of them are non-zero, this results into two 
independent supersymmetry parameters in four dimensions. Therefore equations  
(\ref{a+}-\ref{b-}) describe  an $\N=2$ rather than an $\N=1$ vacuum.     
In order to have $\N=1$ vacua with only NS flux  one of the functions $a$ and $b$ should be  
zero. In this case both pure spinors $\Phi_\pm$ are zero, 
and our equations still hold but do not 
contain much information.
However the supersymmetry conditions for $\N=1$ vacua with only NS 
flux are well known 
\cite{hull,strominger,london}. The structure has to be SU(3), since there
one spinor is involved, and it has to obey
 $(d+H\wedge)(e^{2\phi}\Omega)=0$, 
$e^{2\phi} d(e^{-2\phi} J)=*H$
and $d (e^{2\phi} J^2)=0$. It is not possible to write 
an equation for $\eij$ of the same form as (\ref{a+}).
Nevertheless it is possible to pack all the choices for $a$ and $b$ in a single equation. 
However, in this case the action of $H$ is more complicated than
in (\ref{a+}-\ref{b-}) (see \cite{gmpt}) and at the moment lacks of a 
mathematical meaning, unlike the twist $d+H\wedge$, whose
significance will be reviewed in section \ref{sec:integrability}.
 
\vskip 0.5cm

As we will discuss in detail in section \ref{suff}, 
equations (\ref{a+}),(\ref{a-}) (or (\ref{b+}),(\ref{b-})) together with 
(\ref{eq:norm}) are {\it necessary and sufficient} to find a solution to 
the $\N=1$ supersymmetry conditions. This means that
these equations contain exactly the same amount of information
than the original supersymmetry variations $\delta_{\epsilon}\psi_M=0$ and 
$\delta_{\epsilon}\lambda=0$ with $\epsilon$ given in (\ref{eq:spindec}).
To find a vacuum, one  has to supplement preserved
supersymmetry conditions with Bianchi identities and the equations of motion for the fluxes \cite{Tsimpis}.
In this paper we do not address this issue.

\vskip 0.5cm

Let us now explain briefly how (\ref{a+}-\ref{b-}) were obtained. 
For more details, see \cite{gmpt}, as the method is very similar. 
To set the conventions, we use the string frame 
and the democratic formulation
of \cite{Bergshoeff}. This formulation 
considers all RR fluxes $F_{0,2,4,6,8}$ for IIA and
$F_{1,3,5,7,9}$ for IIB, obeying a duality condition 
$F_n=(-1)^{Int[n/2]} *F_{10-n}$. Since we are interested in 
a maximally symmetric four-dimensional space-time, we only
turn on fluxes that have either none or four components
along it. For a flux with four legs along space-time,
we can use the duality relation to write
it in terms of an internal flux. For example, $F_{\lambda \mu \nu \rho}$
can be traded for $F_6$ fully along internal space. 
As a consequence, all fluxes in (\ref{a+}-\ref{b-}) 
are internal. Secondly, for the supersymmetry
parameters we use the decomposition 
given in (\ref{eq:spindec}).

We start by using the bispinor form of the pure spinors $\Phi_\pm$
given in (\ref{eq:genpure}). This allows to write the exterior derivative
of the \clss\ spinors in terms of the covariant derivative
of the Clifford(6) spinors, namely 
\beq
d\Phi_{\pm}= 
dx^m \wedge \nabla_m \Phi_{\pm}= dx^m \wedge 
\left( (\nabla_m \eta^1_{+}) \otimes \eta_{\pm}^{2 \dagger} 
+ \eta^1_{+} \otimes (\nabla_m \eta_{\pm}^{2\dagger}) \right) \, . \nn
\eeq
The internal component of the supersymmetry 
variation of the gravitino, $\delta \psi_m=0$,
gives the covariant derivative of the spinors 
in terms of the fluxes. 
The equations
are considerably simplified if we use additionally
the dilatino variation $\delta \lambda$ and space-time gravitino variation 
$\delta \psi_{\mu}=0$. The use of dilatino variation is the reason
for the appearance of derivatives
of the dilaton in (\ref{a+}-\ref{b-}), while using the space-time gravitino variation introduces  
the cosmological constant $\mu$. The latter appears 
in the covariant derivative of the supersymmetry parameter along
space-time. To be more precise, there are two possible Killing 
spinor equations for constant negative curvature spaces, namely
$\nabla_\mu \zeta=\frac {\mu_1} 2 \gamma_\mu \zeta$ and 
$\nabla_\mu \zeta=i\frac {\mu_2} 2 \gamma_\mu \gamma_5 \zeta$.
We have kept both of them, as done for example in \cite{ls,bc,Tsimpis}, 
and defined the complex quantity $\mu=\mu_1+i\mu_2$. This complex 
quantity has
the interpretation in four dimensions as a vacuum superpotential \cite{wb}.
The cosmological constant is determined by the norm of $\mu$
as given in (\ref{cc}).

\subsection{Intrinsic torsions for SU(3) and \stt\ structures}
\label{sec:torsions}

In order to prove that the pure spinor equations contain the same information 
as the supersymmetry variations, we have to set up the basic machinery for 
the \stt\ torsions. The way we do this is 
simply to compare it with the better known case of SU(3).

As just said, we start with the case of SU(3) structure, one of the defining 
features of which is the existence of a nowhere vanishing invariant spinor.
In general this spinor is not covariantly constant, but using a bit of SU(3) 
group theory we can write the covariant derivative as
\beq
\label{eq:spjo1}
\nabla_m \eta = i \,  q_m \gamma_7 \, \eta + i \, q_{mn} \gamma^n \eta \, ,
\eeq
where $\gamma_7=-\frac{i}{6!} \epsilon_{mnpqrs} \gamma^{mnpqrs}$.
Equivalently, the invariant forms $J$ and $\Omega$ are not closed 
\begin{equation}
\label{eq:djdo}
\begin{array}{c}\vspace{.3cm} 
dJ = -\frac{3}{2}\, {\rm Im}(W_1 \bar{\Omega}) + W_4 \wedge J + W_3 \, ,\\
d\Omega = W_1 J^2 + W_2 \wedge J + \overline W_5 \wedge \Omega\ \, .
\end{array}
\end{equation}
The covariant derivative of the spinor contains the same information as 
 the exterior derivatives $dJ$ and $d\Omega$ 
\[
(q_m, \, q_{mn}) \leftrightarrow  W_i \, ,
\]
and the precise formulae for this map can be found in \cite{gmpt,fmt}.  

We are ready to give the analogue of these formulae for 
the \stt\ structure. Let us start
by considering the covariant derivative of the spinors. By expanding
$\nabla_m \eta^i_\pm$ in a basis of spinors, we can define the intrinsic torsions:
\begin{equation}
  \label{eq:qi}
\nabla_m \eta^i_+=i q^i_m \eta^i_+ + i q_{mn}^i \gamma^n \eta^i_- \, ,
\end{equation}
which simply duplicates the usual SU(3) structure definition (\ref{eq:spjo1}).
We can use this to compute the exterior derivatives of $\Phi_\pm$, thus
generalizing the $W_i$ of SU(3) structures:
\begin{equation}
  \label{eq:W}
  \begin{array}{c}
  d\Phi_+= W^{10}_m \gamma^m \Phi_+ +W^{01}_m \Phi_+ \gamma^m 
+W^{30}\bar\Phi_-+ W^{21}_{mn}\gamma^m\bar\Phi_-\gamma^n +
W^{12}_{mn} \gamma^m\Phi_-\gamma^n + W^{03} \Phi_- \, ,\\
d\Phi_-= W^{13}_m \gamma^m \Phi_- + W^{02}_m \Phi_- \gamma^m 
+W^{33}\bar\Phi_+ + W^{22}_{mn}\gamma^m\bar\Phi_+ \gamma^n+
W^{11}_{mn} \gamma^m\Phi_+\gamma^n + W^{00} \Phi_+ \, .
  \end{array}
\end{equation}
The change of basis between  (\ref{eq:W}) and (\ref{eq:qi}) can easily be 
found and generalizes the one between (\ref{eq:spjo1}) and (\ref{eq:djdo}); 
the labeling of the components of the 
intrinsic torsion will become clear in a moment.

Since the formal sums of forms are a representation of O$(6,6)$ of which \stt\ 
is  a subgroup, we can decompose the forms under \stt. The decomposition is 
given by the following basis \cite{gualtieri,gualtieri2}, 
analogous to the Hodge diamond:
\begin{equation}
  \label{eq:hodge}
  \begin{array}{c}\vspace{.1cm}
\Phi_+ \\ \vspace{.1cm}
\gamma^m \Phi_+  \hspace{1cm} \Phi_+\gamma^m \\ 
\gamma^m \bar \Phi_- \hspace{1cm} \gamma^m\Phi_+\gamma^n \hspace{1cm} 
\Phi_-\gamma^m\\
\bar \Phi_- \hspace{1cm} \gamma^m\bar \Phi_-\gamma^n \hspace{1cm} \gamma^m
\Phi_-\gamma^n\hspace{1cm}\Phi_-\\
\bar\Phi_-\gamma^m\hspace{1cm} \gamma^m\bar\Phi_+\gamma^n \hspace{1cm} \gamma^m \Phi_-\\
\bar \Phi_+\gamma^m \hspace{1cm} \gamma^m \bar\Phi_+\\
\bar \Phi_+\\
  \end{array}
\end{equation}
Remember that only three of the six $\gamma^m$ 
survive in each of these expressions; for example, since 
$\gamma^m \eta_1^+=\Pi_1^{mn}\gamma_n \eta_1^+$, one has $\gamma^m \Phi_\pm
=\Pi_1^{mn} \gamma_n \Phi_\pm$, where $\Pi_1$ is the holomorphic 
projector with respect to the almost complex structure $J_1$. 
Each entry of this Hodge
diamond should be understood as a representation of \stt: from the top,
$(1,1)$, $(3,1)$, $(1,3)$ and so on.

Returning to  (\ref{eq:W}), we can see now that the superscripts on $W^{ij}$ 
refer to the position of the summand in the Hodge
diamond (\ref{eq:hodge}), with the top element being marked as $00$, the
second row -- $10$, $01$ and so on. 
 We can also notice a few things about the new \stt\ 
intrinsic torsions. First of all, each of them only contains terms
at distance at most three, horizontally or vertically, 
in the Hodge decomposition (\ref{eq:hodge}). This generalizes the fact
that $d\Omega$, for SU(3) structures, contains a form of 
degree $(3,1)$ and (if the almost complex 
structure is not integrable) forms of degree $(2,2)$, but no form of degree 
$(1,3)$. There are additional relations between the $W$'s:
 $W^{30}= W^{33}$, $W^{03}=\overline{W^{00}}$.
These are a generalization of the fact that 
for SU(3) structures, $W_1$ is contained both in $dJ$ and $d\Omega$.
Both these relations follow from the change of basis between (\ref{eq:qi}) and
(\ref{eq:W}).\footnote{They 
can be obtained by deriving the compatibility condition, 
which can be rephrased as 
$\mathrm{Tr}(\Phi^\dagger (\lambda^m)\Phi)=
\mathrm{Tr}(\Phi^\dagger (\iota^m \Phi))=0$ (with $\Phi$ any of the Clifford
vacua in \ref{eq:hodge}) and 
using the constraints of normalization, $\mathrm{Tr}(\Phi_+^\dagger \Phi_+)
 =\mathrm{Tr}(\Phi_-^\dagger \Phi_-)$.}

\subsection{Pure spinor equations vs. supersymmetry}
\label{suff}

We can now finally prove the equivalence of (\ref{a+} - \ref{b-}) and 
(\ref{eq:norm}) to the vanishing of the supersymmetry variations,
for which we will 
use a shorthand ``sufficiency''.
Our strategy is simply to count the number of independent equations
for each \stt\ representation that we get
from the original supersymmetry variations $\delta \psi_M=0=\delta \lambda$
(which take values in Clifford(6)),
and to compare with those obtained  from the set (\ref{a+} -- \ref{b-}) and 
(\ref{eq:norm}) (which are Clifford(6,6)-valued).  


To perform the \stt\ decomposition on both sides 
(Clifford(6) and Clifford(6,6)) we use the fact
that any even or odd real form $F_{\pm}$
can be expanded in the \stt\ basis 
given in (\ref{eq:hodge}) in the following way 
\begin{equation}
  \label{eq:expF}
 F_\pm = F^0\Phi_\pm + F^1_m \gamma^m \bar \Phi_\mp + F_{mn}
\gamma^m \Phi_\pm \gamma^n +F^2_m \Phi_\mp \gamma^m \, \, 
+ \mathrm{( c.~c.)}
\, .
\end{equation}

On the Clifford(6) side, we insert the expansion 
(\ref{eq:expF}) for the NS and RR fluxes 
in the supersymmetry variations $\delta_{\epsilon} \psi_M=0=\delta
_{\epsilon} \lambda$. These equations contain in addition the covariant 
derivative of the internal
spinor, for which we use (\ref{eq:qi}).
This relates $q_m^i$ to $F^i_m$ and $H^i_m$ (the components of
RR and NS flux in their expansion \`a la (\ref{eq:expF}))
and $q_{mn}^i$ to $F_{mn}$,$F^0$,$H_{mn}$ and $H^0$.
We are not going to give the explicit expressions here, as 
they are not particularly enlightening. 
What we are interested in is the number of independent
equations that we get for each \stt\ representation.
There are 
four independent equations for quantities without indices (in the
 (1,1)), eight for quantities with one index only (in the $(3,1)$, $(1,3)$, 
$(\bar 3,1)$ and $(1,\bar 3)$) and four for quantities 
with two indices ($(3,3)$, $(\bar 3, 3)$
and so on).
 
On the Clifford(6,6) side, we plug the decomposition of the RR
and NS fluxes (\ref{eq:expF}) in (\ref{a+}-\ref{b-}).
For the derivative of the pure spinor, we use (\ref{eq:W})
(remember  that there is a one to one correspondence between
$\{W^{ij}\}$ and $(q_m^i,q_{mn}^i)$, so we could use 
the $q$'s for torsions as well). As a result, we get again a set
of equations for each \stt\ representation. After supplementing these
with the equations for the norm of the pure spinors (\ref{eq:norm}),
we confirm that the number of independent equations is the same   
as on the Clifford(6) side (and of course they agree with the latter).

We therefore conclude that (\ref{a+} - \ref{b-}) and (\ref{eq:norm})
are the necessary and sufficient conditions dictated by $\N=1$ supersymmetry
(again, these a priori are not enough to give  solutions to the equations 
of motion since Bianchi identities 
and the equations of motion for the fluxes have still to be imposed).

Note that this procedure gives all the intrinsic torsion components
$W^{ij}$ in terms of fluxes, derivatives of the dilaton 
and warp factor. We do not write the explicit expressions here, but simply 
point out that they are very similar to
the ones given in \cite{gmpt} for the pure SU(3) structure case.

\subsection{Integrability for pure spinors and ${\cal N}=1$ vacua} \label{sec:integrability}

Equations (\ref{a+}) and (\ref{b-}) can be interpreted as an integrability condition for the 
\stt\  structure in IIA and IIB, respectively.


The standard way of introducing an integrability condition involves defining 
{\it  generalized (almost) complex structures} ${\cal J}$. This is like an 
almost complex structure, except that it lives on $T \oplus T^*$ rather than on $T$.
The existence of  ${\cal J}$ reduces the structure group of  \tts to U$(3,3)$. 
Two compatible generalised almost complex structures\footnote{Two 
generalised almost complex structures ${\cal J}_1$ and  ${\cal J}_2$ are compatible if
$[{\cal J}_1, {\cal J}_2] = 0$ and  ${\cal J}_1 {\cal J}_2 = G$, with $G$ a positive definite 
 metric on  \tts.}
reduce the structure group to U(3)$\times$U(3). 
This closely parallels the discussion for pure spinors in the previous section; and 
in fact there is a one to one correspondance between an almost generalized complex 
structure and a pure spinor \cite{hitchin,gualtieri}. Very briefly, consider a single pure 
spinor $\Phi$. By definition, its 
annihilator is a subbundle of \tts of dimension six. It is always possible to find a
generalised almost complex structure that 
has this annihilator as its +i eigenbundle.

The integrability condition then states that the annihilators of the pure spinor 
be invariant under a generalisation of the Lie bracket to $T \oplus T^*$, called the Courant 
bracket. 
Suppose that $A$ is an element of $T \oplus T^*$, a linear combination of the 
$\{ \lambda, \iota \} $ which generate the \clss\ algebra.
It can be considered as an operator on formal sums of forms. 
Given two such objects, we can produce a third one by
\begin{equation}
  \label{eq:courant}
  [A,B]_C(\omega)\equiv d(A B \omega) +A d(B\omega) - Bd(A\omega) -BA 
d(\omega) - (A\leftrightarrow B)  
\end{equation}
where $\omega$ is a differential form. The operation $[\, ,\,]_C$
 is the Courant bracket.\footnote{This definition is more generally known
as {\it derived bracket} \cite{yks}. 
Other particular cases include the Lie bracket 
on vectors, and the Schouten--Nijenhuis on multivectors.} 

Let us see what the integrability condition for the annihilator implies on the pure spinor 
$\Phi$. If we take $\omega$ to be $\Phi$ and $A$ and $B$ both in its 
annihilator, all terms but the last one go away. 
If we take $d\Phi=0$, even the last term goes away and 
$[A,B]_C \Phi=0$ -- that is, $[A,B]_C$ is in the annihilator of $\Phi$. 

So we have shown that $d\Phi=0$ is sufficient for the annihilator of $\Phi$
to be closed under the Courant bracket. (The necessary condition is only
slightly more complicated; $d\Phi$ has to belong to the first Clifford level,
that is, it has to be of the form $C \Phi$ for some $C$ in $T \oplus T^*$.)

Actually, in the definition of the Courant bracket, $d$ could be 
replaced just by any derivation operator. In particular, one can consider 
in (\ref{eq:courant}) the operation $d+ H\wedge$, where $H$ is a closed
3-form, leading to the twisted Courant bracket. If $(d+H \wedge)\Phi=0$,
then the annihilator of $\Phi$ is closed under the twisted Courant bracket.
Therfore, the first equation in (\ref{eq:summa}), 
$(d+H\wedge)\Phi_1=0$, tells us 
that the annihilator of $\Phi_1$ is closed under the twisted Courant 
bracket, or equivalently that the generalized almost complex structure
associated to $\Phi_1$ is integrable. 
Notice that $d+ H\wedge $ this is the 
only combination of $d$ and $H_{mnp}$ 
which is a differential. It has
already been considered in mathematics under the name of twisting 
\cite{hitchin}. 

As we will discuss in Appendix \ref{app:top}, this equation has also appeared in the context of 
topological strings. It is somewhat more curious that also the equation involving RR fields has
a similar structure and involves $(d+H\wedge)\Phi_2$ thus making 
the RR contribution play the role of 
a defect of integrability for the second pure spinor.

\subsection{Complex-symplectic hybrids} \label{sec:local}

We are finally ready to see what kind of manifolds are suitable for type II
compactifications with resulting $\N=1$ vacua. We will set the cosmological
constant $\mu=0$ in this section and for most of the next one, commenting
briefly on the case $\mu\neq 0$ at the end of that section.
As clear from the general form of equations (\ref{a+} - \ref{b-}) 
a twisted closed pure spinor is required and
thus the internal space must be a generalized Calabi-Yau. This still leaves 
plenty of possibilities as far as the differential--geometric structure of the 
internal space is concerned. In order to analyze this, it is useful to return 
to Table 1 and compare the different 
cases from a new perspective. 
For this we will need to introduce the notion of the type of a pure spinor.

In a regular neighborhood of the six-dimensional manifold, 
a pure spinor can be decomposed
in the following way \cite{gualtieri}
\beq
\label{type}
\Phi=e^A \wedge \omega_k \, ,
\eeq
where $A$ is a complex 2-form and $\omega_k$
is a holomorphic $k$-form ($0 \le k \le 3$), which together obey  
$A^{6-2k} \wedge \omega_k \wedge \bar \omega_k \neq 0$. 
The number $k$ is called the {\it type} of the pure spinor. 
It can also be obtained by counting the number of annihilators
which can be expressed as pure $\lambda^m$ -- the intersection of the 
annihilator with $T$. 

The usefulness of this definition is that it gives a local characterization
of a generalized \cy. Thanks of to a generalized Darboux theorem 
\cite{gualtieri}, locally we can always introduce a set of 
holomorphic coordinates $z^1, \ldots, z^k$, augmented by real ones
$x^{2k+1}, \ldots, x^6$: the neighborhood of every point is isomorphic 
to $\bbC^k \times \bbR^{6-2k}$.

Given this definition, we can see from the Table 1 that 
the type of $\eij$ is 0, and the type of $\Omega$ is 3 (this can also be 
obtained using  
the definition of the type in terms of the annihilators). 
Likewise, the types of
the SU(2) pure spinors (\ref{eq:puresu2}) are 2 and 1 for $\Phi_+$ and 
$\Phi_-$ respectively. 
Generically, the type of a pure spinor is as low as allowed by parity. 
This is because imposing that the intersection of the annihilator with the
tangent have a certain dimension is like imposing an equation.
For example, an odd pure spinor will have generically type 1, and may have
type 3 in some loci. So to have a pure spinor of type 3 everywhere, such as
$\Omega$, is very much non--generic. 

The type of generic spinors for \stt\ can be seen by using (\ref{type}) and 
slightly rewriting  (\ref{eq:genpureforms}). We use 
 $j \wedge \omega =\omega^2= 0$ in order to obtain
\[
\Phin_+= 
\frac18 \bar c e^{-i(j+v\wedge w+\frac{1}{\bar c} \omega)}  \ ,
\qquad  
\Phin_-=-\frac18 (v+i w) \wedge e^{i(j+ c \omega)}  \, .
\]
We see that when $c \neq 0$, the first pure spinor is a type 0
(symplectic) spinor, which jumps to type 2 at the points where $c$ 
vanishes. On the other hand,
when $v+iw \neq 0$, the second pure spinor is of type 1, and it jumps to
type 3 at the points where the complex vector vanishes. 
 
Returning to the pure spinor equations (\ref{a+} -- \ref{b-}), again in the 
case $\mu=0$, we see that the integrable spinor for IIA is 
given by $\Phi_+$ and 
for IIB by $\Phi_-$. (Note that parity of the integrable
pure spinor coincides with the parity of the RR fluxes in each theory
-even in IIA, odd in IIB-). 
Thus IIA generally prefers symplectic manifolds, and can 
admit hybrid types when $c=0$, which corresponds to the internal spinors 
being orthogonal; this situation is usually referred to as a static structure.
As for IIB - the situation is reversed, and generally the manifold $M$ is of 
hybrid type (with one complex dimensional part), while at the special 
points of 
``pure'' SU(3) structure, i.~e.~two spinors are proportional and the vector 
vanishes, $M$ has to be complex \footnote{It was already noticed in \cite{Dallagata} 
that a IIB solution on a manifold with SU(2) structure is not necessarily
complex.}.

\section{Discussion}
\label{sec:app}

The necessary conditions for $\N=1$ supersymmetry 
on six manifolds $M$ with SU(3) $\times$ SU(3) 
structure on \tts 
boil down to a pair of equations for two pure spinors:
one tells us that $M$ is a twisted generalized \cy, 
while the second one says that the combined RR fields
act a source for Nijenhuis tensor. The twisted generalized
\cy\ has in IIA an integrable structure that is symplectic
around regular points, and jumps to an hybrid complex--symplectic
(with four and two real dimensions)
at points where the structure is a static SU(2), namely where
the two spinors are orthogonal. In IIB, on the contrary, 
the generalized \cy\ has at regular points an integrable hybrid  
complex--symplectic (with two and four real dimensions).
This integrable structure is purely complex 
at points where the structure becomes pure SU(3), i.e. when the
two spinors are parallel.  

The equations are similar to
those found for topological
models \cite{kapustin,zucchini} and $\N=2$ supersymmetry without
RR fluxes \cite{jw}. 
In the latter case there is a worldsheet description 
with $(2,2)$ supersymmetry \cite{ghr}, and the two pure
spinors are integrable. 
The corresponding schematic equations can be found in the table below:
\begin{center}
\renewcommand{\arraystretch}{1.5}
\begin{tabular}{|c||c|c|c|}\cline{2-4}
\multicolumn{1}{c|}{}
& ${\cal N}=1$ (RR$\neq 0$)& ${\cal N}=2$ (RR$=0$) &top. model\\\hline\hline
\multirow{2}{*}{\stt}&$(d+H\wedge)\Phi_1=0$&
$(d+H\wedge)\Phi_1=0$&\multirow{2}{*}{$(d+H\wedge)\Phi=0$}\\
&$(d+H\wedge)\Phi_2=F$&$(d+H\wedge)\Phi_2=0$&\\\hline
\end{tabular}
\end{center}

\vskip 0.4cm
\noindent
The first entry in this table is the pair of equations obtained in this paper. 
The other two deserve some comments. 

First of all, as we mentioned after (\ref{eq:norm}), the case in which 
$F=0$ and ${\cal N}=1$ 
is not covered by the analysis in this paper; that limit happens to give
${\cal N}=2$ vacua. Indeed, the equations in the second column describe
what is called in \cite{gualtieri} a  ``generalized \cy\ metric'' (stronger than
a generalized Calabi-Yau manifold  \cite{hitchin} used in this paper). 
In \cite{gualtieri}, 
these conditions were found to describe all $(2,2)$ nonlinear 
sigma models: both
the usual \cy\ case and also the  models with twisted multiplets found 
in \cite{ghr}. $(2,2)$ models correspond to ${\cal N}=2$ in the target space,
and indeed the same equations were found to be implied by 
${\cal N}=2$ vacua in supergravity.
It is now also clear why ${\cal N}=1$ with NS only 
could not be described by the $(d+H\wedge)$
closure of two pure spinors: \cite{gualtieri} finds that this would imply 
the existence of a $(2,2)$ sigma model, whereas ${\cal N}=1$ should correspond
to a $(2,1)$ model only. We expand on this in Appendix B.

The last entry in the table is more interesting to us. The condition that 
{\it one} pure spinor be closed is sufficient for a topological model 
to exist on the manifold. This has been argued in \cite{kapustin,zucchini}; 
we will 
say more about it in Appendix \ref{app:top}. 
For now, we can point out that the usual
topological models for SU(3) structure, the A and B models, are obviously 
particular cases: the A model requires the two--form being closed, which 
implies $d\eij=0$; the B model requires the manifold be complex, which is implied
by $d\Omega=0$. 

We would like to point out that the condition for having a 
topological model is, in the Minkowski case, 
one of the two equations we have found, (\ref{a+}) or (\ref{b-}). Hence
in all the Minkowski vacua a topological model can be found.

At this point, however, we should add a word of caution. 
As mentioned many times, we have spelled out the $\N=1$ supersymmetry 
conditions
only, but Bianchi identity is still to be imposed. One of the consequences of
it are the well--known no--go theorem constraining the possibility of 
finding {\it compact} examples.
In the present context, it is easy to see where the problem comes from. 
For Minkowski ($\mu=0$), our equations say essentially $F=d\Phi$. Bianchi
identities together with the flux equations of motion 
say that $F$ is a harmonic form. These two statements together 
imply that $F=0$ on a compact manifold.
Of course there are many noncompact solutions, and they all will fall
in our classification. However  finding interesting compact examples
is clearly of some importance. There are some ways of avoiding the no-go 
theorems which typically involve leaving the supergravity approximation and
including sources or quantum corrections.

These problems are not expected to arise in AdS compactifications. 
While the equations (\ref{a+} -- \ref{b-}) included a cosmological 
constant $\mu$ in the analysis of the equations this was 
set to zero. It is not hard to
see that when $\mu \neq 0$ the generalized \cy\ condition is violated. 
Moreover  we get a mixing between the two pure spinors. In situations 
where $\mu$ is induced by a flux through the four-dimensional spacetime, 
with all
the other  fluxes turned off, we get an extension of the 
nearly-K\"ahler geometry corresponding to pure SU(3) structure, 
which has $d\Omega = \mu J^2$. 
While the case of $\mu \neq 0$ did not receive 
much attention here, it presents readily available examples of
compact manifolds, where in particular Bianchi identities are solved without
introducing orientifolds planes or other complications.

We will finish with two speculations. One practical advantage of having found 
the underlying geometry of $\N=1$ vacua might be a systematic approach to the 
problem of counting moduli. Deformations of generalized \cy\ structures
have already been studied in \cite{gualtieri}. These are yet to be completed 
by 
the second condition involving the RR fluxes, and in principle a coupled system
needs to be analyzed. We may just observe at this point that 
the form of the second equation ((\ref{a-}) for IIA, (\ref{b+}) for IIB) 
suggests that a deformation of $\Phi$ might be simply offset by 
a deformation of the potential $C$. We hope to return on the subject soon.

Finally, these equations might suggest ways to 
find a worldsheet realization of RR backgrounds. This is admittedly a long
shot, but consider the following. First of all, there has been progress 
towards obtaining models with only {\it one} $(d+H\wedge)$ closed 
pure spinor. Second, the fact that we always have a topological model 
associated with $\N=1$ vacua seems to suggest a more profound explanation.
It could be for example that the topological model is related to the 
half--twisting of the $(2,1)$ model that one would expect to be associated
to a ${\cal N}=1$ background.

\vskip 0.5cm
\noindent 
{\bf Acknowledgments}
\vskip 0.3cm

We would like to thank Alberto Cattaneo, 
Marco Gualtieri, Yi Li, Jan Louis, Luca Martucci, Pierre Vanhove, Dan Waldram 
and Maxim Zabzine for useful discussions.
We also thank the Fields Institute and Perimeter Institute for hospitality during the course
of this work, and the organizers and participants of the workshops held there.  
This work  was partially supported by
INTAS grant, 03-51-6346, CNRS PICS 2530,
RTN contracts MRTN-CT-2004-005104 and 
MRTN-CT-2004-503369 and by a European Union Excellence Grant,
MEXT-CT-2003-509661.
MG is supported by European Commission Marie Curie Postdoctoral Fellowship under
contract number MEIF-CT-2003-501485.
AT is supported by DOE contract DEAC03-76SF00515 and by NSF contract 9870115.

\appendix
\section*{Appendices}

\section{The topological model}
\label{app:top}

Here we explain the appearance of the equation $(d+H\wedge)\Phi=0$ in the
topological theory. 

Conventionally, A and B models on a \cy\ are defined as 
topological twistings of a $(2,2)$ 
supersymmetric sigma model. Nevertheless they can be defined directly without any reference 
to twisting and the resulting geometry turns out to be more general.

An intuitive way of thinking about this generalization is an extension of 
the usual properties of A and B models on \cy\ manifolds, where they
depend only on   K\"ahler and  complex structure moduli, respectively.
On a manifold of SU(3) structure one may  think of A and B models
as defined in terms of the pure spinors $\eij$ or $\Omega$, without a priori
imposing both of them to be closed. Given these two examples, it
is natural then to expect that one can associate a topological model 
to any pure spinor.

Let us now see all this more precisely, following \cite{kapustin,zucchini}. 
To define a topological model without twisting a $(2,2)$ model, it is 
essentially enough to give a space of fields and a BRST differential on it. 
The space of fields is made of scalars $\phi^m$ describing the embedding 
of the worldsheet into the target space, and two spinors $\psi^m_\pm$. It is 
customary to define  $T \oplus T^*$ valued $(\rho^m, \chi_m) =(\frac12 (\psi^m_+ + \psi^m_-),
\frac12 g_{mn}(\psi^n_+ - \psi^n_-))$. 

We can then define a candidate BRST differential by $\{Q, \cdot\}$, where 
$\{\, , \}$ is the usual Poisson bracket on the space of fields, and 
\[
Q= ( g_{mn} \partial_0 \phi^n, \partial_1 \phi^m) (1 + i {\cal J})
{\rho^m \choose \chi_m}\ .
\] 
Here $0$ and $1$ refer to the worldsheet indices, and 
${\cal J}$ is a map from $T\oplus T^*$ to itself.
This is the form proposed in \cite{kapustin} for flat space. The general
form has been given in \cite{zucchini} using the formalism of BV 
superfields\footnote{The model can then be written in the usual form by
applying the procedure described in \cite{aksz}. We thank A.~Cattaneo and
M.~Zabzine for discussions on this point.} (see also
\cite{zabzine}); the transformations rules
are similar to the ones given for the $(2,1)$ model in \cite{lmtz}.
Imposing that $Q$ defined this way  gives an honest  differential yields two 
conditions: \\
i) ${\cal J}^2=-1$, where $1$ the identity in $T\oplus
T^*$.\\ ii) the  $i$--eigenbundle of ${\cal J}$, $L_{\cal J}$, is closed under the Courant 
bracket (\ref{eq:courant}). \\
These conditions define a {\it generalized complex structure} \cite{hitchin,gualtieri}.

We may now try to translate this into the language of pure spinors.
The first condition i) implies that the structure group of \tts\
reduces to U$(3,3)$. As shown in \cite{kapustin}, the anomaly cancellation
requires a further reduction of the structure group to 
SU$(3,3)$. This implies that there exists a pure spinor $\Phi$. 
As mentioned in section \ref{sec:integrability}, the integrability of the
generalized complex structure, i.e. condition ii)
translates into $d\Phi=0$. Introduction of $H$--flux (the twisting) modifies this to 
$(d + H\wedge) \Phi=0$.  

We conclude by putting together this general situation with the particular cases 
of the A and B models:
\begin{center}
\begin{tabular}{|c||c|c|c|}\hline
model & structure & $Q^2=0$ & anomaly\\\hline\hline
A & $J_{mn}$ & $dJ=0$ & ---\\\hline
B& $J_m\,^{n} $ & $J_m\,^{n} $ integrable & $c_1=0$\\\hline
generalized & ${\cal J}$ & ${\cal J}$ integrable & $\exists$ pure spinor\\\hline
\end{tabular}  
\end{center}

\section{Courant bracket and purely NS ${\cal N}=1$ vacua}
\label{app:N1}

In this appendix we briefly discuss the integrability properties of
the ${\cal N}=1$ backgrounds with only NS fields (restricted to the case
of SU(3) structure) \cite{hull,strominger}. As mentioned in the text,
these do not satisfy two pure spinor equations of the form
(\ref{eq:summa}).

If we denote the space of annihilators of a pure spinor $\Phi$ as
$L_{\Phi}$,
we may introduce further subspaces of $L$ in the following way. Since the
two pure spinors
are compatible, $T \oplus T^*$ splits into four subspaces, each
annihilating
one of the four corners of the Hodge diamond (\ref{eq:hodge}). Then in an
obvious notation, we call these $L_\nearrow$, $L_\nwarrow$,
$L_\searrow$,
$L_\swarrow$. In models with $(2,2)$ worldsheet supersymmetry ($\N=2$
spacetime) all four subspaces are closed.

The $\N=1$ NS backgrounds have pure SU(3) structure and 
$(2,1)$ worldsheet supersymmetry.
The relevant equations are
\begin{equation}
  \label{eq:N1}
 i\partial J = H^{2,1} \ , \qquad 
d(e^{2\phi} J^2)=0\ , \qquad (d+H\wedge)(e^{2\phi}\Omega)=0.  
\end{equation}
where $H^{2,1}$ is the (2,1) component of $H$. 
It follows that the manifolds in question have to be 
complex, and that by scaling the metric, we may
define a closed holomorphic three-form. Thus $\Omega=\Phi_-$ is a
generalized \cy\ structure, and therefore the space of its annihilators
$L_{\Omega} = L_\nearrow \oplus L_\searrow$ is closed under the twisted
Courant bracket (we are using the version of (\ref{eq:courant})  with the
differential being $d + H\wedge$). On the contrary, $\eij$ is not a
$(d+H\wedge)$--closed and as a consequence, $L_\nearrow \oplus L_\nwarrow$
is not integrable.

However, let us consider $L_\nearrow = L_{\Omega} \cap  L_{e^{iJ}}$. It is
generated by elements of the
form $\iota_{\partial_i} - g_{i\bar j} dz^{\bar j}$. An easy
computation
shows that these are closed under the twisted Courant bracket if and only
if the first equation in (\ref{eq:N1}) is satisfied.

One may check that $L_{\Omega} = L_\nearrow \oplus L_\searrow$ and
$L_\nearrow = L_{\Omega} \cap L_{e^{iJ}}$ are the only pieces closed under
twisted Courant bracket, thus leaving us with fewer integrable sectors
than for $\N=2$, yet still preserving the generalized \cy\ structure.

\end{document}